\documentclass[aps,superscriptaddress,showpacs,twocolumn]{revtex4}

\usepackage{graphicx,graphics,epsfig}   
\usepackage{dcolumn}    
\usepackage{bm}         
\usepackage{amsmath}    
\usepackage{verbatim}   
\usepackage{color}      
\usepackage{times}
\usepackage{amsmath,amsfonts,amssymb,graphics,graphics,color,times}

\usepackage{latexsym}
\usepackage{amsmath}
\usepackage{amssymb}
\usepackage{amsfonts}
\usepackage{amsthm}
\usepackage{mathrsfs}
\usepackage{color,verbatim,graphics}
\DeclareMathAlphabet{\mathrsfs}{U}{rsfs}{m}{n}
\DeclareMathAlphabet{\mathpzc}{OT1}{pzc}{m}{it}
\DeclareMathAlphabet{\matheus}{U}{eus}{m}{n}
\DeclareMathAlphabet{\mathbbold}{U}{bbold}{m}{n}

\newcommand{\ba}{\begin{eqnarray}}
\newcommand{\ea}{\end{eqnarray}}
\newcommand{\ban}{\begin{eqnarray*}}
\newcommand{\ean}{\end{eqnarray*}}
\newcommand{\be}{\begin{equation}}
\newcommand{\ee}{\end{equation}}

\newcommand{\ket}[1]{|#1\rangle}
\newcommand{\bra}[1]{\langle#1|}

\begin{document}

\title{Proposal for a loophole-free Bell test based on spin-photon interactions in cavities}

\author{Nicolas Brunner}
\affiliation{H.H. Wills Physics Laboratory, University of Bristol, Tyndall Avenue, Bristol, BS8 1TL, United Kingdom}
\affiliation{D\'epartement de Physique Th\'eorique, Universit\'e de Gen\`eve, 1211 Gen\`eve, Switzerland}
\author{Andrew B. Young}
\affiliation{Centre for Communications Research, Department of Electrical and Electronic Engineering, University of Bristol, Merchant Venturers Building, Woodland Road, Bristol, BS8 1UB, UK}
\author{Chengyong Hu}
\affiliation{Centre for Communications Research, Department of Electrical and Electronic Engineering, University of Bristol, Merchant Venturers Building, Woodland Road, Bristol, BS8 1UB, UK}
\author{John G. Rarity}
\affiliation{Centre for Communications Research, Department of Electrical and Electronic Engineering, University of Bristol, Merchant Venturers Building, Woodland Road, Bristol, BS8 1UB, UK}

\begin{abstract}
We present a scheme to demonstrate loophole-free Bell inequality violation where the entanglement between photon pairs is transferred to solid state (spin) qubits mediated by cavity QED interactions. As this transfer can be achieved in a heralded way, our scheme is basically insensitive to losses on the channel, and works also in the weak coupling regime. We consider potential experimental realisations using single atom, colour centre and quantum dot cavity systems. Finally our scheme appears to be promising for implementing quantum information protocols based on nonlocality. Here we discuss a possible implementation of device-independent quantum key distribution.
\end{abstract}

\maketitle

\section{Introduction}

The correlations between separated observers performing local measurements on an entangled state cannot be explained by a classical mechanism. A signal is impossible, as it would have to travel faster than light. Pre-established correlations are ruled out via the violation of a Bell inequality \cite{bell}. This phenomenon termed quantum nonlocality is at the very heart of quantum mechanics and arguably one of the most surprising and counterintuitive aspects of the theory (see \cite{review} for a recent review). In recent years it has received growing attention, partly due to the development of device-independent quantum information processing \cite{bhk,DI,DI2,randomness,randomness2,vazirani1,vazirani}, which is based on quantum nonlocality. The idea is that the correct implementation of a protocol can be guaranteed without resorting to assumptions about the devices used in the protocol. For instance, in device-independent quantum key distribution \cite{DI}, two distant parties can establish a secret key and guarantee its privacy, via a Bell inequality violation, without placing assumptions about the detailed functioning of their devices.
Notably there is hope that this approach will be able to address some of the shortcomings of standard QKD schemes \cite{makarov}. Its practical implementation is however still challenging \cite{gisin,curty}.

Quantum nonlocality has been demonstrated in numerous experiments. However, technical imperfections in these experiments open loopholes, which makes it still possible, in principle, to explain the data through a classical mechanism. Given the fundamental significance of quantum nonlocality, it is highly desirable to perform a Bell experiment free of any loopholes. 
Realizing such a test is challenging as it requires (i) a space-like separation between the parties, and (ii) a high enough detection efficiency.
Failure to address (i) opens the locality loophole: the correlations can be explained by a sub-luminal signal. Failure to address (ii) opens the detection loophole \cite{pearle}: a classical model exploiting the detector inefficiency can fake Bell inequality violations. 
Experiments carried out on photons could achieve (i) while atomic experiments could achieve (ii). However no experiment was yet able to close both loopholes simultaneously. In particular, typical photon detection efficiencies are still too low to close the detection loophole. 
Significant progress was nevertheless achieved in the last years, both from the experimental \cite{harald,Wittmann,Giustina,brad} and theoretical \cite{CI,CV1,CV2,Asym,Qudits,CS} point of view.

These issues must also be addressed for the implementation of device-independent protocols. Here loopholes (in particular the detection loophole) could in principle be exploited by the eavesdropper, and hence compromise the device-independent security of the protocol \cite{DI2}. Therefore the perspective of a loophole-free Bell experiment, and more specifically of achieving detection-loophole-free Bell violations on long distances, is essential for implementing protocols such as device-independent QKD.

Here we present a scheme for a loophole-free Bell test based on Faraday-type spin-photon interactions in cavities \cite{HR08,DK04,HR09}. Our setup is hybrid, using both photons and 'atomic' systems, hence combining the best of both worlds. Using spin-photon interactions, the entanglement between two photons can be swapped to two atomic systems in cavities. Importantly this can be realized in a heralded way via the detection of the photons after interaction with the cavities, even in the weak coupling regime. Once the creation of an entangled pair is heralded, the measurement settings for the Bell test are generated. 
In this way the setup, being 'event-ready', is basically insensitive to the losses in the optical channel between Alice and Bob. Moreover, since spin measurements have efficiency close to unity, the setup is immune to the detection loophole. These features make our scheme particularly well adapted to the implementation of device-independent quantum key distribution. In the following we start by presenting a simple theoretical model of our setup, estimating the effect of technical imperfections such as the amplitude of the spin-photon interaction and the decay rate of the atomic system in the cavity. Then, we discuss the implementation of our scheme in various platforms, including negatively charged nitrogen vacancy centres (NV$^-$) in diamond, quantum dot spin systems and atoms. Finally, we discuss the implementation of device-independent QKD based on our setup.

\section{Setup}

The setup is sketched in Fig.~1. A pair of polarization entangled photons is emitted by a central source and sent to two remote observers, Alice and Bob. The photons are in the singlet state
\ba \ket{\psi_-}_{AB} = \frac{1}{\sqrt{2}} (\ket{R,L}-\ket{L,R}) \ea
where we have used the notation $ \ket{R,L} = \ket{R}_A \otimes \ket{L}_B$, and $\ket{L}$ ($\ket{R}$) denote a left (right) circular polarization. At each observers laboratory, the photon is then sent through an optical cavity containing a spin 1/2 particle. The photon and spin couple via a Faraday type effect, whereby the polarization of the photon is rotated depending on the state of the spin. Here this interaction is represented by a unitary operation of the form
\ba U(\alpha) = e^{i \alpha ( \ket{L}\bra{L} \otimes \ket{\uparrow}\bra{\uparrow} + \ket{R}\bra{R} \otimes \ket{\downarrow}\bra{\downarrow})} \ea
where $\ket{\uparrow}$ and $\ket{\downarrow}$ represent the two orthogonal spin states. The parameter $\alpha$ represents the strength of the interaction. Experimentally, the achievable values of $\alpha$ depend on the physical system that is considered, as we will discuss below.

Initially, both spins are prepared in the superposition state $(\ket{\uparrow}+\ket{\downarrow})/\sqrt{2}$. Hence the initial state of the global system, photons and spins, is given by
\ba \ket{\Psi} = \ket{\psi_-}_{AB} \otimes \frac{\ket{\uparrow}_A+\ket{\downarrow}_A}{\sqrt{2}} \otimes \frac{\ket{\uparrow}_B+\ket{\downarrow}_B}{\sqrt{2}} \ea
After each photon interacted with the spin in the cavity, the state becomes
\ba && U(\alpha_A)  U(\alpha_B) \ket{\Psi} = \\ 
&& \ket{R,L} (e^{i\alpha_B} \ket{\uparrow\uparrow} +  \ket{\uparrow\downarrow}   e^{i(\alpha_A+\alpha_B)} \ket{\downarrow\uparrow} + e^{i \alpha_A} \ket{\downarrow\downarrow} ) 
   \nonumber \\  \nonumber  && - \ket{L,R} (e^{i\alpha_A} \ket{\uparrow\uparrow}  
    + e^{i(\alpha_A+\alpha_B)}  \ket{\uparrow\downarrow} + \ket{\downarrow\uparrow} + e^{i \alpha_B} \ket{\downarrow\downarrow} ) 
\label{unitary}
\ea
where we have omitted subscripts and normalization.

\begin{figure}[t!]
\begin{center}
  \includegraphics[width=\columnwidth]{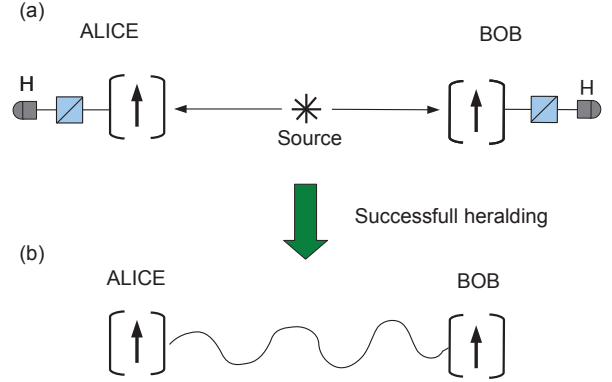}
  \caption{Setup of the experiments. (a) A pair of polarization entangled photons is sent to two remote observers. Each photon interacts with a spin inside a cavity. (b) Upon successful heralding, i.e. both photons detected as horizontally polarized at the cavity's output, the two spins are prepared in an entangled state, and the Bell test can be performed.}
\label{fig1}
\end{center}
\end{figure}

After exiting the cavity each photon is measured in the horizontal-vertical polarization basis. For simplicity we will focus here on the case in which both photons are detected in the horizontal mode, i.e. projected onto the state $\ket{H} = \frac{1}{\sqrt{2}}(\ket{L}+\ket{R})$. The final state of the two spins (here unormalized) is then found to be
\ba \ket{\Psi_f} &=& (e^{i\alpha_A}-e^{i\alpha_B}) \ket{\phi_-} + (e^{i(\alpha_A+\alpha_B)}-1) \ket{\psi_-}
\ea
where 
\ba \nonumber \ket{\phi_-}=\frac{1}{\sqrt{2}} ( \ket{\uparrow \uparrow}- \ket{\downarrow\downarrow})\, \, , \quad \ket{\psi_-}= \frac{1}{\sqrt{2}}(\ket{\uparrow \downarrow}- \ket{\downarrow\uparrow}) \ea
are the usual two-qubit Bell states.
The state can now be conveniently rewritten as
\ba \label{psif} \ket{\Psi_f} = \frac{1}{\sqrt{N}} \left( \sin(\Delta \alpha) \ket{\phi_-} + \sin(\bar{\alpha} \right)
\ket{\psi_-}) \ea
where $\Delta \alpha = (\alpha_A-\alpha_B)/2$ and $\bar{ \alpha} = (\alpha_A+\alpha_B)/2$ are the difference and average values of the spin-photon interactions (of Alice and Bob) respectively, and $N= \sin^2(\Delta \alpha)+\sin^2(\bar{\alpha})$ is a normalization factor. The state \eqref{psif} is entangled, unless $\Delta \alpha = \bar{\alpha}$. Note that the probability of entangling the two spins, i.e. to detect both photons in the horizontal polarization mode, is given by $N/4$. Upon heralding of the spin entanglement, the measurement settings for the Bell test are generated. This operation takes an amount of time $t$ during which the spins decohere. Here we model this effect via an exponential decay. After a time $t$, the state of the spins becomes 
\ba \label{rhof} \rho_f = e^{-t/\tau} \ket{\Psi}\bra{\Psi} + (1-e^{-t/\tau}) \frac{\mathbb{I}}{4}\ea
where $\mathbb{I}/4$ is the fully mixed state of two qubits.

\section{Testing the CHSH Bell inequality}

Now that we have characterized the state of the two spins, we will investigate its ability to violate a Bell inequality. Indeed our goal is to define the range of parameters that will lead to nonlocal correlations. 
Here we will focus on the Clauser-Horne-Shimony-Holt (CHSH) Bell inequality \cite{chsh}. In this case, both Alice and Bob can perform two possible binary measurements. The choice of measurement settings of Alice (Bob) is denoted $x=0,1$ ($y=0,1$), and the outcome $a=\pm1$ ($b=\pm1$).
For a pair of measurement settings, $x,y$, the correlation between the measurement outcomes of Alice and Bob is given by
\ba E_{xy} = p(a=b|x,y) - p(a \neq b|x,y) \ea
The CHSH expression is then given by
\ba \text{CHSH} = E_{00} + E_{01} + E_{10} - E_{11} \leq 2, \ea
which holds for any local model.
Violation of this inequality hence witnesses nonlocality.

\begin{figure}[t!]
\begin{center}
  \includegraphics[width=\columnwidth]{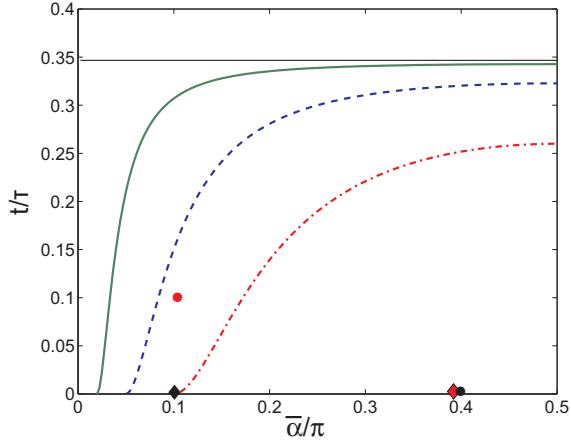}
  \caption{Requirements for violation of the CHSH Bell inequality. Each curve represents a set of parameters such that CHSH=2; hence in the region below each curve the CHSH inequality is violated. Here the data is shown for different values of $\Delta \alpha$, the difference (or error) in the spin-photon interaction; the thin horizontal line is $\Delta \alpha=0$, the solid green curve is $\Delta \alpha=\pi/50$, the dashed blue curve is $\Delta \alpha=\pi/20$, and the dash-dotted red curve is $\Delta \alpha=\pi/10$. The points are estimated values taking present experimentally achievable parameters: semiconductor quantum dots (red dot), NV$^-$ centers in diamond in high/low-Q cavities (black/red diamond) and atom cavity systems (black dot). For details see the experimental realisations section}
\label{fig2}
\end{center}
\end{figure}

For our final state \eqref{rhof}, we now determine what set of parameters (i.e. strength of spin-photon interaction, spin decoherence) in order to obtain a violation of the CHSH inequality. Note first that in the (unrealistic) case in which both spin-photon interaction can be set to be perfectly equal, i.e. $\alpha_A=\alpha_B>0$, the state \eqref{rhof} takes the particularly simple form
\ba \rho_f = e^{-t/\tau} \ket{\psi_-}\bra{\psi_-} + (1-e^{-t/\tau}) \frac{\mathbb{I}}{4}\ea
that is, a two-qubit Bell state mixed with white noise. Thus, Alice and Bob should use the optimal CHSH measurements, tailored for the singlet state $\ket{\psi_-}$. In this case the CHSH inequality is violated whenever $e^{-t/\tau}>1/\sqrt{2}$, i.e. $t/\tau < \ln{\sqrt{2}} \simeq 0.3466$. Importantly, CHSH violation occurs here regardless of the strength of the spin-photon interaction, i.e. for any $\alpha_A=\alpha_B>0$. 

In the more realistic case in which $\Delta \alpha$ is non-zero, the state \eqref{rhof} takes a more complicated form. Luckily, the maximal CHSH violation of any two-qubit state can be evaluated using the criterion of Ref. \cite{horo}. The results are presented in Fig.~2. Notably we observe now a trade-off between mean interaction strength $\bar{\alpha}$ and the spin decoherence $t/\tau$. In particular, and as expected intuitively, for weak interaction quick spin measurements are demanded, whereas for stronger interactions slower spin measurements can be tolerated. 
In the following we will evaluate the feasibility of our setup, how stringent the requirement of Fig.~2 are, for various physical platforms.

\section{Experimental realisations}

The realisation of the unitary in equation (2) involves a degenerate polarisation-dependent lambda transition strongly coupled to a cavity. The polarisation dependent transitions arise in atomic transitions with spin half ground states \cite{RR12}, certain transitions in NV$^-$ centers \cite{JW04} and charged quantum dots \cite{YR11}. Here we specialise to a reflection geometry because it can be shown \cite{HR08} that reflected probe light will experience a polarisation dependent phase shift $\alpha$ (and consequent Faraday rotation) which becomes large when cavity mode volume is small and the cavity storage time measured by its quality factor $Q$ is high. The reflected photon is thus correlated with the ground state spin and coincident detection of two horizontal polarised photons can be used to herald entanglement of two remote spins. 

We estimate the achievable $\bar{\alpha}$ assuming a (lossy) single sided cavity containing a single two level system (atom) in the weak excitation regime \cite{HR08, DK04}. The reflection coefficient and phase shift $\alpha$ at frequency $\omega$ is given by 
\begin{eqnarray}\label{eqn:ref}
r(\omega)&=&|r(\omega)|e^{i\alpha} \\ \nonumber &=& 1-\frac{\kappa(i(\omega_{d}-\omega)+\frac{\gamma}{2})}{(i(\omega_{d}-\omega)+\frac{\gamma}{2})(i(\omega_{c}-\omega)+\frac{\kappa}{2}+\frac{\kappa_{s}}{2})+g^{2}}\nonumber
\end{eqnarray} 
where $g$ is coupling rate between the `atom' and the cavity, $\gamma$ is `atom' decay rate, $\omega_c , \omega_d$, are the cavity and `atom' resonances respectively and $\kappa$ is the decay rate from the cavity $\kappa_{s}$ the decay rate due to losses with $\kappa+\kappa_{s}=\omega_c/Q$. 

For demonstration of loophole-free Bell inequality violation we require that the spin is stored long enough for the herald photon events to be registered as a coincidence detection which sets a minimum value of measurement delay $ t\sim D/c +T_l$ where $D$ is the separation of observers and $T_l$ is a minimum detection and electronic logic delay (of order 100ns). Once a coincidence event is recorded, the observers generate locally their measurement settings. We also require the measurement time uncertainty, the time taken to manipulate and read out the spin state $\Delta t$, to be small compared to $D/c$ to avoid the locality loophole. Thus we can summarise the requirements for a loophole free test as 
\begin{equation} 
\label{D}
\Delta t<D/c \ll\tau 
\end{equation}

The read out of a spin cavity system that is strongly coupled involves the measurement of the Faraday rotation induced by the spin on an incident linear polarised beam. Using diagonally polarised probe light and measuring reflected light in a polarization beamsplitter oriented to differentiate horizontal ($H$) and vertical ($V$) polarizations the spin state can be estimated from the difference in photon counts detected in the $H$ and $V$ detectors $N_H-N_V=\pm N_{tot}\sin{\alpha/2}$ with total detected photocounts $N_H+N_V=N_{tot}$ and with positive (negative) outcome signaling spin up (down). The read out time is limited by the constraint of having $<0.1$ photon on average in the cavity at any one time and having a reasonable certainty in the spin measurement which is guaranteed when $\sin \bar{\alpha} \gg 1/\sqrt{ N_{tot}}$. This constrains measurement time to be 
\begin{equation} 
\label{readout}
\Delta t \gg 10\tau_c/\sin^2 \bar{\alpha}   
\end{equation}
where $\tau_c=Q/\omega_c$ is the cavity lifetime. By suitably enforcing inequality \eqref{readout}, we can ensure the fidelity of the readout
is near unity, this implies near unit efficiency (hence ensuring that the detection loophole is closed).

For the case of NV$^-$ and atomic transitions the ground states are often separated by an energy such that only one state is in resonance with the cavity at a time. Equation \eqref{unitary} is simplified to $U(\alpha) = e^{i \alpha ( \ket{L}\bra{L} \otimes \ket{\uparrow}\bra{\uparrow} )} $ but the underlying maths does not change if we halve $\bar{\alpha}$ and take this into account in the following calculations. 

In the case of NV$^-$ manufacturing high Q low loss cavities is also difficult. We could exploit recently fabricated low Q-factor microcavities \cite{MB12} in order to perform a loophole free Bell test. Here using very small cavity volumes the interaction ($g$) between the NV$^-$ and the light can be greatly enhanced while the cavity Q-factors remain low ($\kappa \ll g$) and strong coupling is not reached. Instead we operate in the one-dimensional atom regime \cite{PhysRevA.75.053823} where by suitably adjusting the cavity input-output coupling efficiency and looking close to resonance with the NV$^-$ zero phonon line we can achieve a large phase shift approaching $\pi$ albeit with lowered reflectivity (see Appendix). 
However when we move from strong coupling into the weak coupling regime the minimum readout time is now determined by the lifetime of the state $\tau_s$ rather than that of the cavity $\tau_c$ in equation \eqref{readout}. In such small cavities this can be reduced by an order of magnitude from bulk lifetimes due to Purcell enhancement.  The readout time is also extended by the reduced cavity reflectivity in the weak coupling regime.

We summarise state of the art for the four systems in Table 1 estimating parameters from the published work on strongly coupled atom \cite{RR12}, NV$^-$ \cite{PW06} and dot \cite{YR11} cavity systems and the low-Q photonic crystal cavities \cite{MB12}. For simplicity we have estimated $t$ from twice the minimum readout time (assuming equality in \eqref{D}). The results show that the NV$^-$ and atom cavity systems satisfy the requirement that the delay between storage and readout should be very short compared to decoherence time $\tau$ . However the long cavity storage times needed to achieve strong coupling in these systems mean that the minimum separation of measurement stations is of order 150 $m$ for the atoms and 100 $m$ for NV$^-$. At first sight the  measurement time on the dot cavity system is also short enough to mean that decoherence is small, however we have to take into account the inevitable electronic delays which mean the minimum $t\sim 100 ns$ and thus we expect $t/\tau \sim 0.1$. The low-Q cavity results are very encouraging as cavities with these specifications are already being fabricated \cite{MB12}. We estimate that the reflectivity of these systems will still be $>30\%$ and thus counting losses will not be severe. However we note that for all realisations the linewidth of the entangled pair photon source must be smaller than the cavity linewidth (or transition linewidth in the case of low-Q cavities) which will require sophisticated filtering if standard crystal sources are to be used and pair emission rates will be limited to $R\simeq 0.1/\tau_c$ ($\tau_s$ for low-Q case) which is hundreds of thousands per second for atoms and NV$^-$ and many million per second for dot systems. We have plotted the nominal positions of the various realisations on figure 2 and can clearly see that loophole free violations can be expected when $\Delta \alpha<\pi/10$. Hence the experiment could in principle be carried out in the weak coupling regime, although one should indeed ensure that the probability of a successful heralding is large compared to photo-detector dark counts.

Finally, we give an estimate of the total time required to perform the experiment. As an example, we consider the case of a low-Q cavity with $\bar{\alpha} = 0.4 \pi$ and $\Delta \alpha = \pi/10$. The expected Bell violation is then $\text{CHSH} \simeq 2.32$. Considering a separation of $300m$ between Alice and Bob and a transmission loss of $3 dB/km$ (considering photons at wavelength $ 700 nm$) we obtain a channel transmission of $\eta_t \sim 80\%$. The probability of a successful heralding is then $p_{herald} = \eta_t  \eta_c^2  \eta_d^2 \frac{N}{4} \sim 7 * 10^{-4}$ per photon pair emitted from the source, where $\eta_c$ is the reflectivity of the cavity and $\eta_d$ is the efficiency of the photo-detectors (here we take $\eta_c = \eta_d = 30 \%$). To match the cavity, a narrow-band (cavity enhanced) source of entangled photons is required \cite{rob}, which can produce $\sim 10^6$ pairs per second. We thus expect $\sim 10^3$ heralding events per second. Hence the main limitation comes from the measuring time, here estimated to be of order $ms$. Good statistics, say $10^5$ runs, would then necessitate a few minutes of data acquisition.

\begin{table}
\begin{center}
\begin{tabular}{|c||c|c|c|c|}
\hline
System & Atoms & N-V centres & Dots & Low-Q \\
\hline \hline
$g/(2\pi)$ (MHz) & 5  & 100  & 5000 & 3300\\ 
\hline
$\kappa/(2\pi)$ (MHz) &  3 & ~13 & ~3000 & 440000\\
\hline
$\kappa_s/(2\pi)$ (MHz) &  ~0.5 & ~39 & ~7000 & 220000\\
\hline
$\gamma/(2\pi)$ (MHz) & 3 & 0.6 & 1000 & 6\\
\hline
$\bar\alpha$ (Rad) & $\sim 0.4\pi$ & $\sim 0.1\pi$ & $\sim 0.1\pi$ & $\sim 0.4\pi$ \\
\hline
$\tau$   ($\mu$s) &  10000 &    1000 &    1  & 1000\\
\hline
$\Delta t$ ($ns$) & 500 &  300 & 1.5 & 1000\\
\hline
Min $D$  (m) &  150  & 100 & $<1$ & 300\\     
\hline
Min $t$  ($\mu$s) & $\sim 1$ & $\sim 0.3$ & $\sim 0.1$ & $\sim 2$ \\
\hline
\end{tabular}
\caption{Estimated parameters for the implementation of our setup in various physical platforms.}
\end{center}
\end{table}

\section{Device-independent QKD}    

The setup discussed here could also be adapted for the implementation of device-independent (DI) protocols, such as DI-QKD. Two possibilities can be considered. First the scheme of Fig.~1, featuring a spin-photon interface on both sides, and a source of photons placed in the middle. Note however that the requirements for implementing DI-QKD are slightly different compared to the case of a loophole-free Bell test (see discussion in \cite{DI2}). In particular, it is important to ensure that no information can leak out of Alice and Bob's labs, for instance by an adequate shielding. Thus, the setup will be immune against the locality loophole, and the timing constraints discussed in the previous section can be relaxed. Here the main requirement will be that $\Delta t < \tau$. Another possibility consists in using the spin-photon interface only on Bob's side, and placing the source close to Alice's lab. Then the spin-photon interface is used to herald the presence of a successful transmission of a signal to Bob's lab, similarly to the `qubit amplifier' of Ref. \cite{gisin}. 

\begin{figure}[b!]
\begin{center}
  \includegraphics[width=\columnwidth]{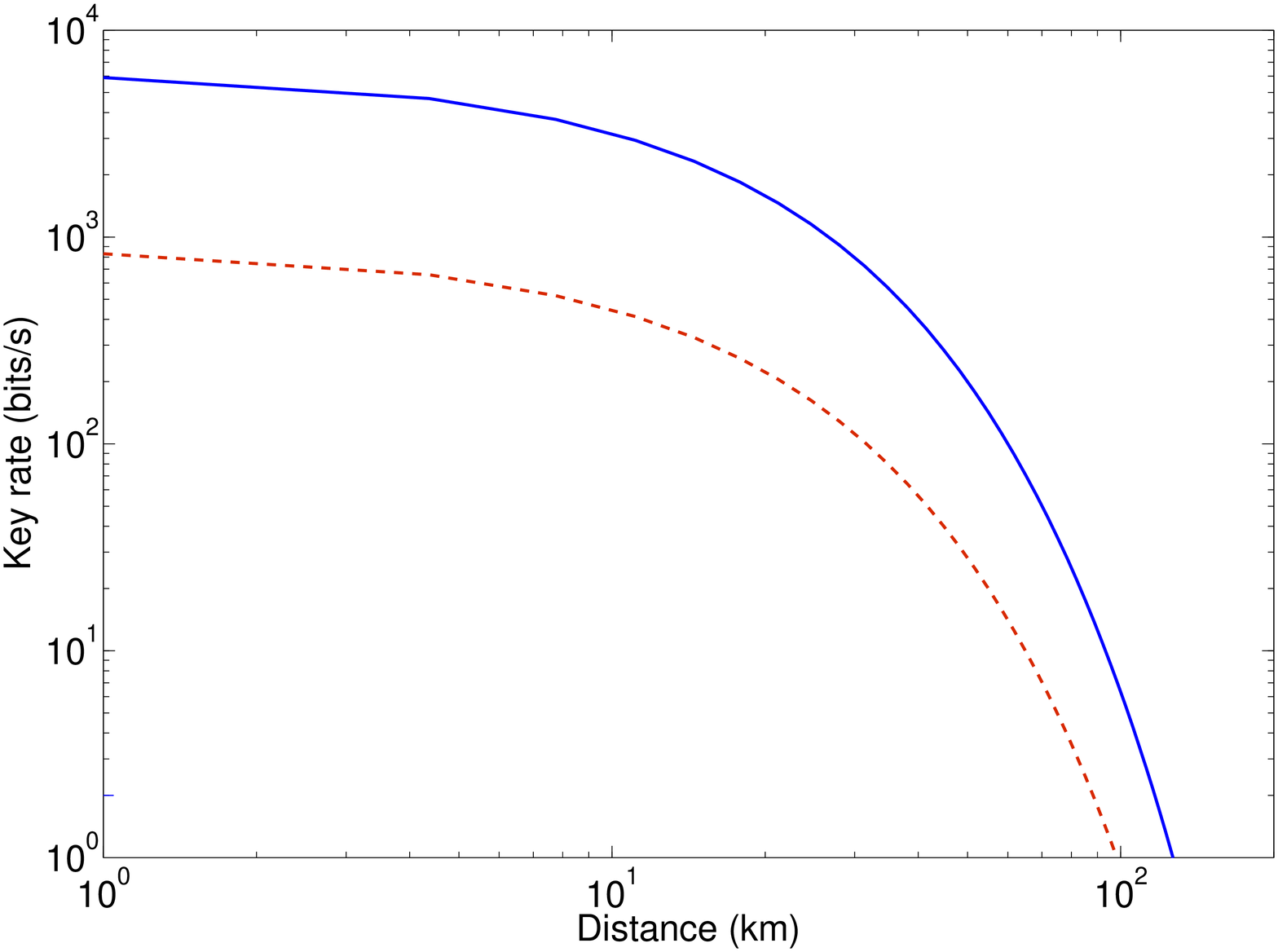}
  \caption{Key rate as a function of the distance for an implementation of DI-QKD using spin-photon interface. The present analysis considers an implementation based on quantum dots. We have considered two sets of parameters for the cavity's reflectivity $\eta_c$ and the photo-detection efficiency $\eta_d$: $\eta_c=0.3$ and $\eta_d=0.5$ (dashed red line), and $\eta_c=0.5$ and $\eta_d=0.8$ (solid blue line). }
\label{fig2}
\end{center}
\end{figure}

Finally, we give an estimate of the key rate following the first approach mentioned above (i.e. the scheme of Fig.~1). As we are considering much larger distances here compared to the case of a loophole-free Bell test, it will be important to choose a setup in which the wavelength of photons is in the telecom window to limit losses. Here, as an example, we consider an implementation based on quantum dots working with photons at $\lambda = 1.3 \mu m $, with Faraday interactions given by $\bar{\alpha} = 0.1 \pi$ and $\Delta \alpha = \pi/50$. The expected Bell violation is then $\text{CHSH} \simeq 2.45$. We consider a source working at $10 GHz$ \cite{zhang}, with a probability of pair creation of $10^{-3}$, and expect transmission losses of $0.3 dB/km$. Following Ref. \cite{DI2}, which considers security against collective attacks, the key rate is given by
\ba R = 1- h(q) -h(\frac{1+ \sqrt{(CHSH/2)^2-1}}{2}) \ea
where $h(x)= -x \log_2{x} -(1-x) \log_2 (1-x)$ is the binary entropy. Here $h(q)$ represents the fraction of the raw key that Alice and Bob must sacrifice in order to correct for errors; note that here the error is given by $q= \sin^2{\Delta \alpha}/(\sin^2{ \bar{\alpha}} + \sin^2{\Delta \alpha}) \simeq 4 \%$. In Fig.3, we plot the key rate as a function of the distance. We see that for distances up to $100km$ we get decent key rates, considering reasonable parameters for the reflectivity of the cavity and the photo-detector efficiency. We believe that this opens promising perspectives for implementing DI-QKD using the present scheme. A detailed study of the performance of the spin-photon interface (and its implementation in different physical platforms) would be of great interest \footnote{Note that Mattar and coworkers \cite{mattar} recently discussed in detail the implementation of DI-QKD using spin-photon systems, and reported promising results, in particular high key rates.}.

\section{Conclusion}    

We discussed a hybrid scheme where entanglement between photons is transferred to solid state (spin) qubits mediated by cavity QED. The successful entanglement of the spins can be heralded by the detection of the photons. The spins can be measured efficiently within times that allow both the locality and the detection loopholes to be closed. The required apparatus separations are measured in hundreds of metres when we consider atom and NV$^-$ centres ground states as our matter qubits in realisable cavities while this could be reduced to metres if the electron spin of singly charged quantum dots is used as the qubit. In this latter case the remaining challenge is to achieve electron spin decoherence times longer than 1 $\mu s$ possibly through sophisticated spin echo schemes \cite{spin}. Finally, we discussed potential applications of these systems for the implementation of device-independent protocols.

\emph{Note added.} While finishing this manuscript, we became aware of a related work by Sangouard and colleagues \cite{sangouard} who proposed a loophole-free Bell test based on heralded mapping of photonic entanglement into single atoms. Also, while the present work was under review, Mattar and colleagues discussed in detail the implementation of DI QKD using the present setup. 

\emph{Acknowledgements.}
We thank A. Acin and C. Simon for discussions. This work is supported by the UK EPSRC, the EU projects DIQIP and Q-ESSENCE, the Swiss National Science Foundation (grant PP00P2\_138917) and ERC Advanced Grant (QUOWSS 247462).


\section{Low Q-factor cavity for loophole free Bell test}

Here we show that a low q-factor photonic crystal microcavities fabricated in single crystal diamond around a single $NV^{-}$ centre [21] could be suitable for performing a loophole free Bell test. The ratios of $g:\kappa_{T}:\gamma$ are estimated from [21], and can be found in Table 1 in the main document. We then calculate the reflected amplitude and phase when photons are reflected from the $NV^{-}$ cavity system using equation \eqref{eqn:ref} defining $\kappa_{T}=\kappa+\kappa_{s}$ as the total decay rate from the cavity setting the Q-factor. 

\begin{figure}[t!]
\begin{center}
\includegraphics[width=\columnwidth]{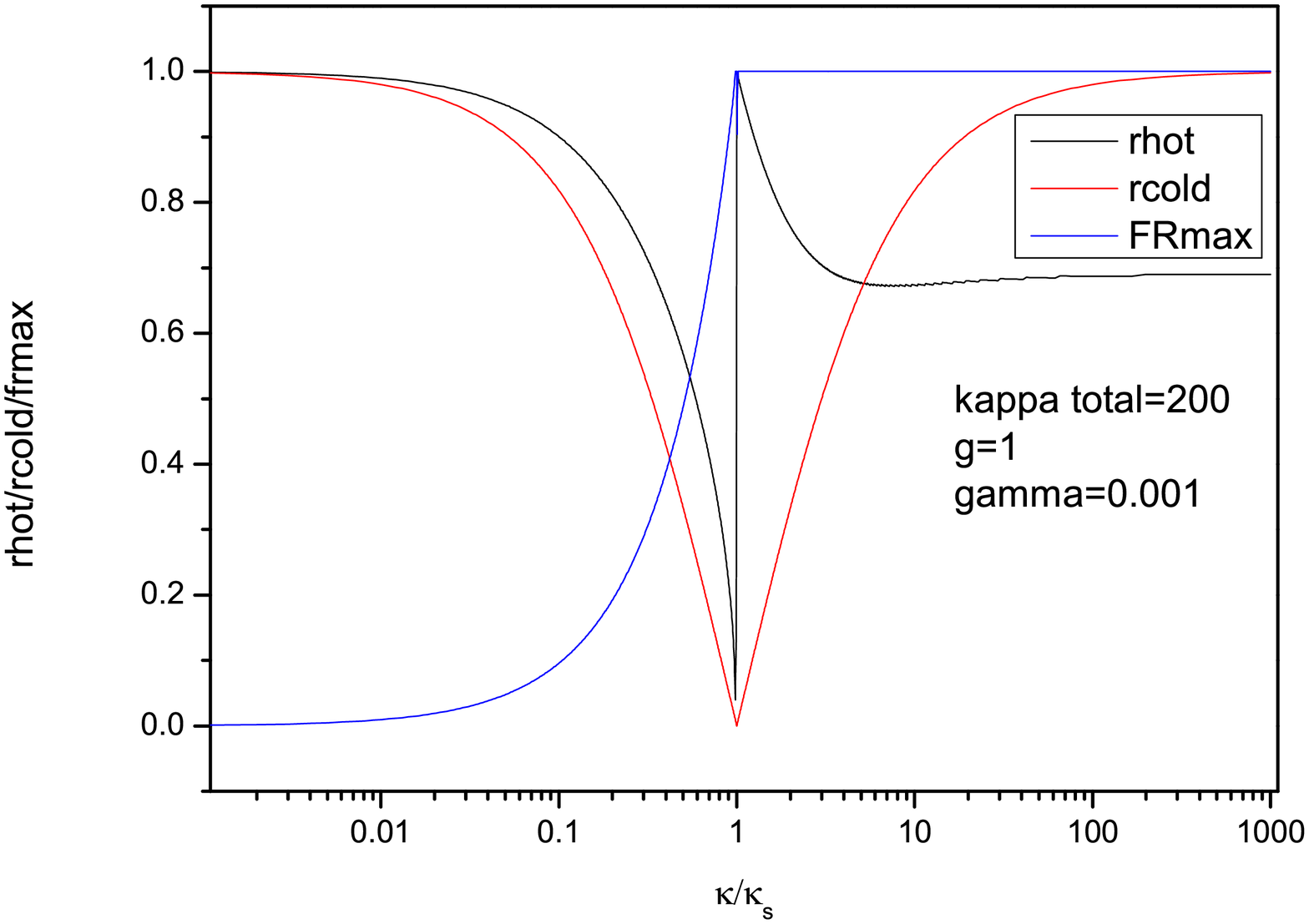}
\caption{Showing the reflectivity for an $NV^{-}$ centre coupled to a photonic crystal cavity using experimental parameters from [21]. We then examine the change the the maximum faraday rotation angle as a function of the ratio $\kappa:\kappa_{s}$ whilst keeping the Q-factor constant and the same as in [21].  $FR_{max}$ represents the sine of the maximum Faraday rotation. $r_{hot}$ represents the reflectivity from the case when the NV is coupled to the cavity and $r_{cold}$ represents the reflectivity from an empty cavity, at the point where the faraday rotation is a maximum.} 
\end{center}
\end{figure}

The calculated reflectivity and phase can now be seen in Fig.1. where we have kept $\kappa_{T}$ constant and varied the ratio of $\kappa:\kappa_{s}$. Ideally for optimum efficiency we would like to work in the loss free regime where $\kappa>>\kappa_{s}$ however this is not a realistic goal in present day microcavity structures. A more realistic target is to work in the regime where $\kappa\approx4\kappa_{s}$, this has the added benefit as at this point the empty cavity reflectivity ($r_{cold}$) is approximately equal to cavity with resonant $NV^{-}$ centre ($r_{hot}$) so we do not need to modify the equations in the main document. At this point the maximum Faraday rotation is $\alpha= \pi/2$ ($\sin(\alpha)=1$) and $r_{hot}=r_{cold}=0.65$ which leads to a high fidelity protocol with a decrease in efficiency.

It is worth pointing out that whilst for simplicity here we have considered the case where $r_{hot}=r_{cold}$, the fidelity of the preparation of the $\ket{\psi^{-}}$ state we are interested in is unaffected by non-equal reflectivity's between hot and cold cavities, and differences between the reflectivity of Alice and Bob's respective spin photon-interfaces.

\end{document}